\documentclass[times,twocolumn,final]{elsarticle}
\usepackage{sigcon}
\usepackage{framed,multirow}

\usepackage{amssymb}
\usepackage{latexsym}

\usepackage{url}
\usepackage{xcolor}

\usepackage{hyperref}

\definecolor{newcolor}{rgb}{.8,.349,.1}

\usepackage{booktabs}
\usepackage[numbers]{natbib}
\usepackage{amssymb}
\usepackage{graphicx}
\usepackage{tikz}
\usepackage{dblfloatfix}
\usepackage{caption}
\usepackage{subcaption}
\usepackage{xcolor}
\usepackage{multirow}
\usepackage{array}
\usepackage{wrapfig}
\usepackage{amsmath}
\usepackage{color, colortbl}
\definecolor{Gray}{gray}{0.925}

\usepackage[T1]{fontenc}
\usetikzlibrary{positioning,decorations.markings,shapes.symbols,calc,backgrounds,shadows} 

\usetikzlibrary{shapes,arrows}
\usetikzlibrary{automata,positioning}
\usepackage{dsfont}
\usepackage{pgfplots}
\pgfplotsset{width = 6cm, compat = newest}
\usepackage{sansmath}
\pgfplotsset{
  tick label style = { font=\tiny\sansmath\sffamily},
  label style      = { font=\small\sansmath\sffamily}
}
\usepackage{color,soul}

\mathchardef\period=\mathcode`.
\DeclareMathSymbol{.}{\mathord}{letters}{"3B}

\journal{Biomedical Signal Processing and Control }

\begin{document}

\let\WriteBookmarks\relax
\def\floatpagepagefraction{1}
\def\textpagefraction{.001}

\verso{H. A. Qadir \textit{et~al.}}

\begin{frontmatter}

\title{Simple U-net Based Synthetic Polyp Image Generation: Polyp to Negative and Negative to Polyp}%

\author[1]{Hemin Ali Qadir}

\author[1,2]{Ilangko Balasingham}
\author[3]{Younghak Shin\corref{cor1}}
\cortext[cor1]{Corresponding author: 
  email: shinyh0919@gmail.com}

\address[1]{The Intervention Centre, Oslo University Hospital (OUS), Sognsvannsveien 20, 0372 Oslo, Norway}
\address[2]{Department of Electronic Systems, Norwegian University of Science and Technology (NTNU), Høgskoleringen 1, 7491 Trondheim, Norway}
\address[3]{Department of Computer Engineering, Mokpo National University, Jeollanam-do, Muan-gun, Cheonggye-myeon, Yeongsan-ro, 1666 KR, South Korea}

\received{}
\finalform{}
\accepted{}
\availableonline{}
\communicated{}

\begin{abstract}
%%%
Synthetic polyp generation is a good alternative to overcome the privacy problem of medical data and the lack of various polyp samples. In this study, we propose a deep learning-based polyp image generation framework that generates synthetic polyp images that are similar to real ones. We suggest a framework that converts a given polyp image into a negative image (image without a polyp) using a simple conditional GAN architecture and then converts the negative image into a new-looking polyp image using the same network. In addition, by using the controllable polyp masks, polyps with various characteristics can be generated from one input condition. The generated polyp images can be used directly as training images for polyp detection and segmentation without additional labeling. To quantitatively assess the quality of generated synthetic polyps, we use public polyp image and video datasets combined with the generated synthetic images to examine the performance improvement of several detection and segmentation models. Experimental results show that we obtain performance gains when the generated polyp images are added to the training set.
%%%%
\end{abstract}


\end{frontmatter}

%\linenumbers

% Main text
\section{Introduction}\label{intro}
Colorectal cancer (CRC) is the third most frequent cause of cancer mortality and the second leading cause of death in men and women globally \cite{bray2018global}. Polyps are precursors to colorectal cancer, and resection of polyps can effectively prevent the development of colorectal cancer \cite{zauber2012colonoscopic}. Colonoscopy is an important standard tool for colon screening globally. It has the advantage of being able to visualize and remove lesions simultaneously during screening. However, the effectiveness of a colonoscopy depends on the skills and expertise of the endoscopy specialist. Therefore, the polyp detection rate varies widely, and the polyp miss rate is known to be between 6 and 27\% in clinical trials \cite{ahn2012miss}. 

Over the past two decades, many researchers have been working on automatic polyp detection using efficient computer-aided detection systems. Early studies attempted to detect polyp by extracting features such as color, shape, and texture of polyp \cite{karkanis2003computer, tajbakhsh2016convolutional, shin2018automatic}. Recently, attempts have been made to automatically detect and segment polyps using deep learning models which improved performance over traditional feature extraction-based methods \cite{bernal2017comparative, yu2016integrating, shin2018automatic2, qadir2019improving, qadir2019polyp, qadir2021toward}.

Most deep learning-based studies use several public polyp datasets \cite{bernal2017comparative, shin2018automatic2, qadir2019improving, qadir2019polyp, qadir2021toward} for research purposes and focus on developing deep learning model architectures with proper pre- and post-processing methods to improve polyp detection and segmentation performance. One of the difficulties in polyp detection and segmentation is the diversity of their features. Polyps vary widely in color, shape, size, and texture. However, widely used polyp datasets only have dozens of unique polyps used for deep learning-based detection and segmentation model training \cite{bernal2015wm, silva2014toward, angermann2017towards}. Because in medical fields, abnormal data is inherently more difficult to collect than normal data. To obtain the location information required for polyp detection and segmentation, manual annotations by colonoscopy experts are essential and are usually costly and logistically difficult.

In order to overcome the limitation of availability of large polyp image data, many deep learning-based studies are actively using transfer learning and image augmentation techniques \cite{shin2018automatic2, tajbakhsh2016convolutional}. However, to effectively train a deep learning model with a large number of learnable parameters, it is necessary to increase the diversity of polyp appearances within the training dataset.

Recently, a GAN(generative adversarial network)-based image generation approach has also been applied to polyp data to overcome insufficient training datasets \cite{mahmood2018unsupervised, shin2018abnormal}. In \cite{mahmood2018unsupervised}, GAN based domain adaptation approach has been applied to the Endoscopy dataset. They focused on adapting real data to synthetic-like data to remove patient-specific details from real images. This is the difference from our study, which aims to generate new synthetic polyp images to improve polyp detection and segmentation performance. 

In \cite{de2021training}, to generate a synthetic polyp, a polyp mask is first generated using a GAN framework, and the polyp part is generated when the polyp mask is given using the conditional GAN approach. Then, the polyp-free part is found in the real polyp image and the generated polyp is synthesized. The difference from the proposed approach is that two GAN frameworks should be used for polyp generation. In addition, a separate algorithm for each image is required to find the polyp-free part, and another separate algorithm is used to naturally synthesize the generated polyp part into the existing image. On the other hand, the proposed method uses a single conditional GAN framework to convert a polyp image into a negative image and convert the negative image back to a polyp image. Therefore, there is no need to use a separate algorithm to find a polyp-free part or perform natural synthesis.

In \cite{shin2018abnormal}, a conditional GAN was used to generate synthetic polyps. The authors proposed to use a conditional mask to control the position, size, and shape of generated polyps and showed improved polyps detection performance when the generated polyps were added as additional training data. However, negative images must be used to generate the polyp image, and additional processing (i.e., edge filtering) was required to create the input condition image. Furthermore, it is not possible to generate polyps with various characteristics from the same input condition. From a performance verification point of view, the performance improvement was weakly evaluated, i.e., only one deep learning model was evaluated for the polyp detection task.

In this paper, to overcome the shortcomings of the previous polyp image generation work, we conduct a study that generates synthetic polyp images similar to real ones using only a given polyp dataset without the use of additional datasets and processes. We propose a framework that uses the same GAN architectures to convert the polyp images into a negative image and the negative image back into a novel polyp image. In addition, we attempt to generate various polyp features by introducing a controllable polyp mask with one input condition image. The experimental results show performance improvement when synthetic polyp images generated by the proposed method are used for training various deep learning models for object detection and segmentation.

The remainder of this paper is organized as follows. In Section \ref{method}, the proposed polyp to negative and negative to polyp image generation framework including network architecture and preparation of input conditioned images are introduced. We also explain the polyp detection and segmentation models used for the qualitative evaluation of generated polyp images. In Section \ref{expermint_data}, experimental datasets used in this study are described. In Section \ref{results}, experimental results and discussions are described. Finally, we conclude this study in Section \ref{conclusion}. 

\begin{figure*}[t]
    \centering
    \includegraphics[trim=5cm 7.35cm 10cm 1.9cm, clip=true, scale=0.8]{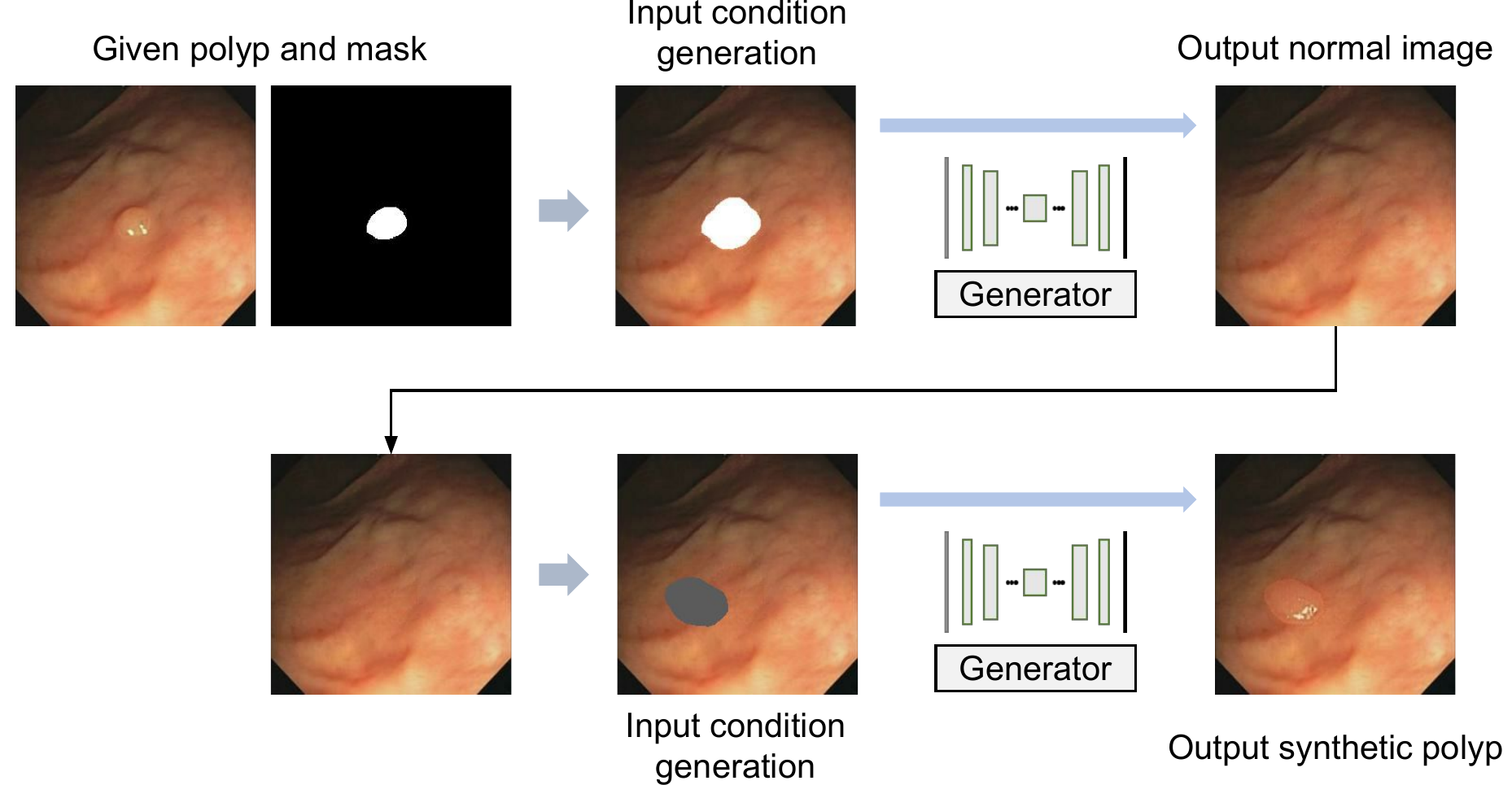}
    \caption{Proposed conditional GANs based polyp image generation framework.}
    \label{fig:fig_1}
    \vspace{-4mm}
\end{figure*}

\begin{figure*}[!b]
    \centering
    \includegraphics[trim=5.5cm 5.65cm 8cm 3.575cm, clip=true, scale=0.6]{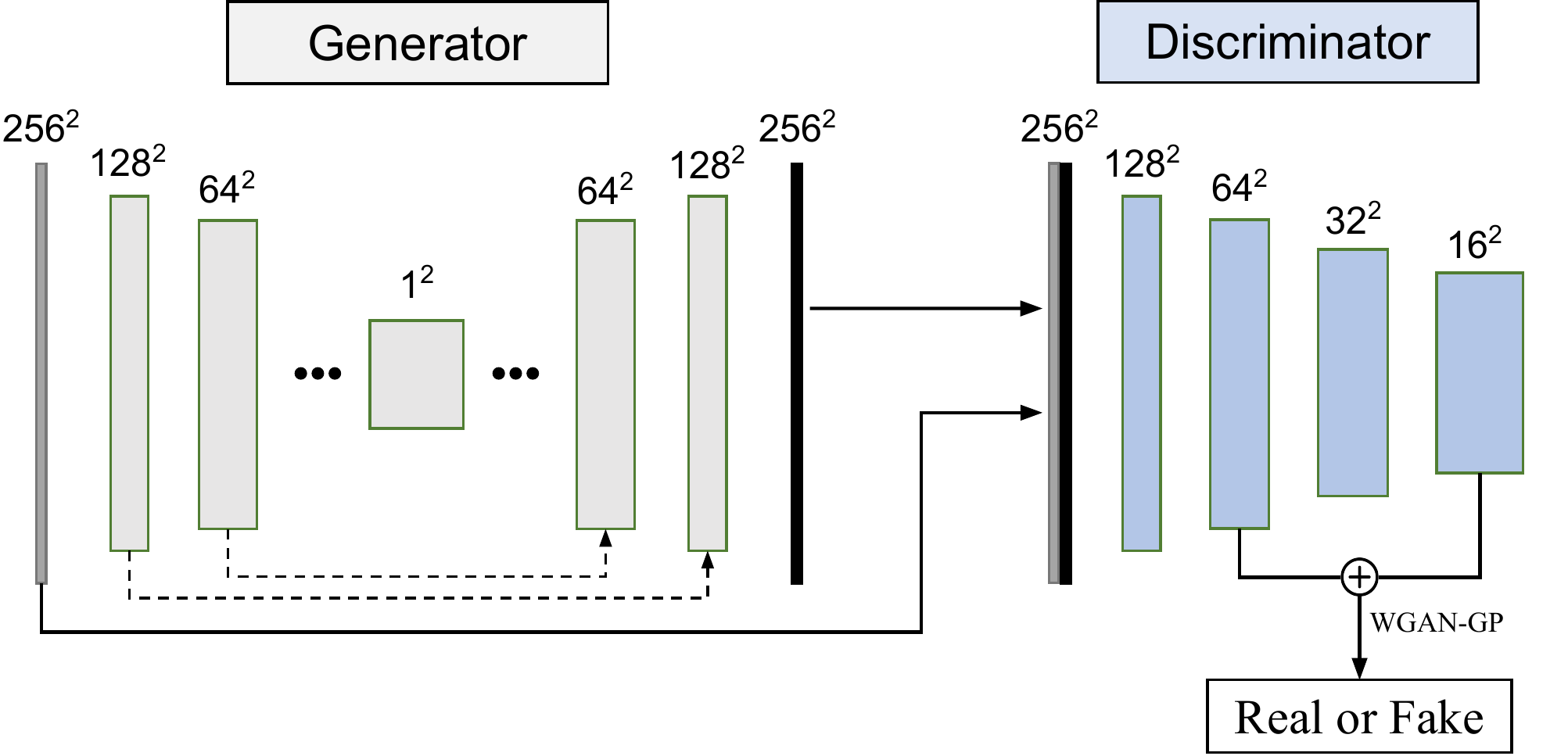}
    \caption{Proposed U-Net based generator and discriminator architectures. The same architecture is used for both polyp to negative and negative to polyp tasks.}
    \label{fig:fig_2}
    \vspace{-2mm}
\end{figure*}

\section{Methods}\label{method}
Figure \ref{fig:fig_1} shows the concept of the proposed synthetic polyp generation procedure. Using a given polyp image and its corresponding label mask, we first transform the polyp part to the negative part. Then, we transform the generated negative image into a new synthetic polyp image using a controllable input condition mask. The controllable mask is used to force the generator network to produce polyps with different characteristics depending on the color of the polyp mask in the input condition image (see Section \ref{Normal2Polyp}). For both transformations, we use the same network architectures. We explain each method in the following subsections in detail.

\subsection{Network Architectures}\label{network}
Figure \ref{fig:fig_2} shows the structure of the generator and discriminator networks used for the conditional GAN framework in this study. Recent GAN studies have focused on proposing new network architectures to obtain task-specific high performance image generation \cite{shin2018abnormal}, image inpainting \cite{yuan2019image} and image detection and segmentation \cite{qadir2021toward, shvets2018angiodysplasia}. On the other hand, we adopt a simple U-net based generator architecture used in \cite{isola2017image} to make it universally work for two different tasks (e.g., converting from polyp to negative and from negative to polyp). The difference from the paper \cite{isola2017image} is that a multi-patch discriminator is used in the discriminator and WGAN-GP loss \cite{gulrajani2017improved} is adopted as the GAN loss function. Detailed explanations are provided in the following paragraphs.

As shown in Figure \ref{fig:fig_2}, for the input image size of 256×256, 4×4 convolution with stride 2 is applied in each encoding layer, so that the feature map size is reduced to 1/2 and then progresses to the final 1×1 feature map. In each subsequent decoding layer, transposed convolution is applied to increase the feature map size by 2 and a final 256×256 image is generated. In the encoder part of the generator, we use Convolution-BatchNorm-leakyReLU module in each layer while in the decoder part, we use transposed Convolution-BatchNorm-ReLU module. Dropout is used in the first 3 layers of the decoder part with a dropout rate of 50\%.

In the case of the discriminator, a patch-based discriminator which is widely used for image-to-image transformation tasks is applied \cite{shin2018abnormal, isola2017image}. Additionally, to reflect the diversity of polyp sizes well, a multi-patch discriminator was used that utilizes two output patches at the same time as shown in Figure \ref{fig:fig_2}. In the discriminator, the Convolution-BatchNorm-leakyReLU module was applied to the input image in the same way as the encoding layer of the generator, and the final feature map size is 16×16 as shown in Figure \ref{fig:fig_2}. For the multi-patch discriminator, GAN loss for 64×64 and 16×16 feature maps is added.

Our training loss is comprised of a GAN term and a reconstruction (Reconst) term which are commonly used in the existing methods \cite{shin2018abnormal, isola2017image}: \begin{equation}
     arg \: \underset{G}{min} \: \underset{D}{max} \: L_{GAN}(G. D) + \lambda L_{Reconst}(G).
\end{equation} where $G$ refers to the generator network, $D$ refers to the discriminator network, and $\lambda$ is a parameter to control the balance between two different loss terms. In the case of the GAN loss term, we use the WGAN-GP loss \cite{gulrajani2017improved}, which is known to improve training stability and has been recently used in many GAN studies \cite{gulrajani2017improved, shaham2019singan}. For reconstruction loss, we use the L1 loss between the generated image and the original target image \cite{isola2017image}. L1 loss is needed to model low-frequency structure (blurry results) because the GAN loss can only model high-frequency structure and give much sharper results. The framework can introduce virtual artifacts when the GAN loss is used alone \cite{isola2017image}. When both losses are used together, they lead the framework to generate realistic images to fool $D$.

%relying on an L1 term to force low-frequency correctness 

\subsection{Polyp Image to Negative Image Generation}\label{Polyp2Normal}
To generate a new polyp image, we first train our conditional GAN model to transform a given polyp image to a new negative image. To do this, we need to prepare a pair of training images, i.e., input conditioned image and output target image as shown in Figure \ref{fig:fig_1}. 

\begin{figure}[!ht]
    \centering
    \includegraphics[scale=0.45]{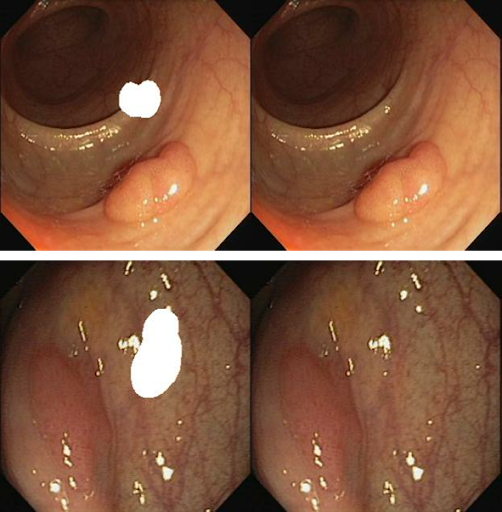}
    \caption{Training image pairs for polyp to negative image generation. Left: input condition image, Right: output target image.}
    \label{fig:fig_3}
    \vspace{-2.5mm}
\end{figure}

Each row in Figure \ref{fig:fig_3} is an example of a pair of images used for training polyp to negative image transformation. An input condition image (shown on the left side in Figure \ref{fig:fig_3}) is created from combining a polyp training image and a generated mask. The original polyp image (shown on the right side in Figure \ref{fig:fig_3}) is used as an output target image. Thus, the proposed conditional GAN trains an image inpainting task that targets generating the negative parts in the right images by using the masks in the left images in Figure \ref{fig:fig_3} as its condition. 

To create the masks shown in the left images in Figure \ref{fig:fig_3}, we apply basic image augmentation such as rotation, scaling, position translation, and perspective transform with randomly selected parameters to the original polyp masks in the training dataset. Since the network will be trained to reconstruct the negative image portion only, the generated masks are randomly placed in the input condition images, avoiding overlapping with the polyp portion.

\begin{figure}[!ht]
    \centering
    \includegraphics[scale=0.2]{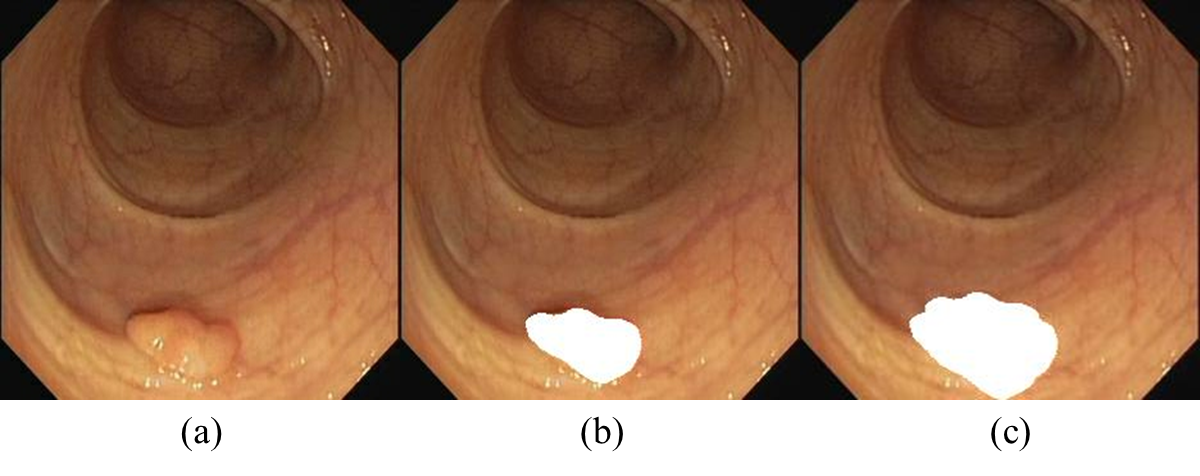}
    \vspace{-3mm}
    \caption{Example of input conditioned image for negative image generation in inference. (a) original polyp image (b) original polyp image combined with original polyp mask (c) original polyp image combined with extended polyp mask.}
    \label{fig:fig_4}
    \vspace{-2mm}
\end{figure}

In the inference phase, the goal is to transform the polyp region in a given polyp image to a negative region. The polyp image is combined with its corresponding polyp mask to create the input condition image. However, if we use the original polyp mask, shadows or connections around the polyp will remain as shown in Figure \ref{fig:fig_4}(b). This shadow will appear as an artifact and  prevent generating a realistic negative image. To avoid this issue, we dilate the original polyp mask by 10 pixels (as shown in Figure \ref{fig:fig_4}(c),) as the final mask for the input condition in the inference.

\subsection{Negative Image to Polyp Image Generation}\label{Normal2Polyp}
\begin{figure}[!ht]
    \centering
    \includegraphics[scale=0.5]{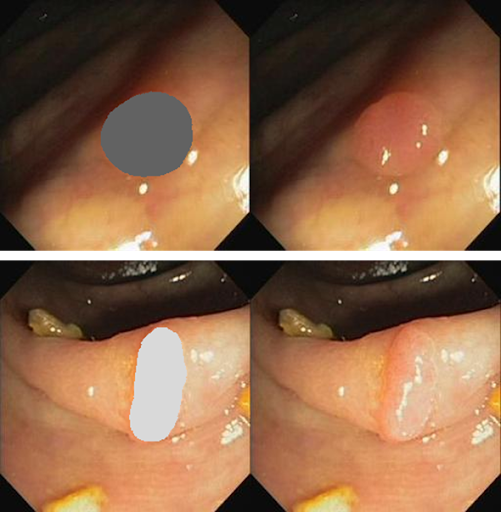}
    \caption{Training image pairs for negative to polyps image generation. Left: input condition images, Right: output target images (original polyp images).}
    \label{fig:fig_5}
    \vspace{-3mm}
\end{figure}

Next, we prepare training data for the final objective, the negative image to polyp image transformation task. Figure \ref{fig:fig_5} shows two pairs of training images used for the negative image to polyp image transformation. The input condition image is generated from combining the polyp images with their corresponding polyp masks provided with the datasets. Figure \ref{fig:fig_5} shows two pairs of training images: the input condition image is shown on the left side, and the output target shown on the right is the polyp image. We aim to train the generator network to regenerate the same form of polyp in the given mask region. 

In the previous image to image translation method \cite{isola2017image}, there is a shortcoming in that the same image is always generated for the same input condition. Also, in \cite{shin2018abnormal}, there is a shortcoming in that the characteristics of the generated polyps for various input conditions are very similar. The reason is that a single binary polyp mask was used for various polyps. To overcome this problem, we try to use a controllable input condition. It is intended to generate a new polyp by adjusting the mask values even if the the shape of the mask in the input condition is the same shape (see the left side in Figure \ref{fig:fig_5}).

The dataset of polyp images used for training comprises 34 different unique polyps (see the information of CVC-ClinicDB dataset in section \ref{expermint_data}). We allocate 34 different values evenly distributed between 0 and 255 rather than traditional binary masks for the polyp regions in the training images. Therefore, the value of the polyp masks varies depending on the polyp in the given image as shown on the left side of Figure \ref{fig:fig_5}, meaning a different gray-scale value is assigned for each unique polyp in the dataset.

By controlling the value of the polyp masks in the input condition images, we can generate a polyp with different features and characteristics from the same input condition of the same shape at the inference phase, see Section \ref{results} for detailed generation results. When inferring the final polyp generation, the input condition image is produced by combining a label mask, which is created from a randomly generated shape and selected value, with a negative image, which is obtained by the polyp to negative image transformation model.

\subsection{Polyp Detection and Segmentation Models}\label{Det_Seg}
For quantitative evaluation of the generated polyp images, we assess the performance of several polyp detection and segmentation models when trained with and without generated synthetic polyps. In more detail, performance evaluation is performed on several detection and segmentation models trained only with the data given as the original training set and the models trained with the generated polyp images added to the original training set.

For polyp detection, we select three well-known detection models: Faster R-CNN ResNet101, Faster R-CNN Inception Resnet v2 \cite{ren2015faster} and  R-FCN ResNet101 \cite{dai2016r}. Faster R-CNN utilizes a pre-trained deep CNN as a feature extractor and uses the last convolution layer as a region proposal network (RPN) \cite{ren2015faster}. Then, the classification and regression are performed on the RoIs (Region of Interest) proposed by the RPN. We use two deep CNN models i.e. ResNet101 and Inception Resnet v2, as the feature extractors pre-trained on Microsoft's (MS) Common Objects in Context (COCO) datasets.

R-FCN has been proposed to effectively overcome the slow training and inference speeds occurring during RoI-specific processing of the existing Faster R-CNN \cite{dai2016r}. R-FCN implements the position-sensitive score maps technique that can include information about the location of objects while sharing the operation of RoIs extracted from RPN using a fully convolutional layer. Resnet101 pre-trained on MS COCO dataset is used as the feature extractor of R-FCN.

We also use three segmentation models to qualitatively evaluate the generated synthetic polyps when used as additional training data. All three models are based on encoder-decoder architecture of the U-Net family: TernausNet-16 \cite{shvets2018angiodysplasia, iglovikov2018ternausnet}, AlbuNet-34 \cite{shvets2018angiodysplasia}, and MDeNetplus \cite{qadir2021toward}. U-Net is a widely used segmentation model in the field of medical imaging where training data is limited \cite{ronneberger2015u}. A U-Net architecture consists of an encoder-decoder structure with skip connections. The skip connections enable feature combinations of the encoder layers and decoder layers, which allows for sophisticated localization. The TernausNet-16 model uses ImageNet pre-trained VGG-16 for the encoder network, while the AlbuNet-34 uses ImageNet pre-trained ResNet-34 as an encoder and improves the method of skip-connections from U-Net \cite{shvets2018angiodysplasia}. MDeNetplus has not only skip connections from the encoder layers to decoder layers, but also feedback connections. The feedback connections sum the activation maps of similar layers of different decoders.

\subsection{Model Training}\label{Training}
We train the GAN network from scratch and the Adam optimizer with a momentum of 0.5 at a learning rate of 0.0002. Before training, the input images of 256×256 are resized to 312×312 and then randomly cropped back to 256×256 for applying random jittering \cite{isola2017image}. For other parameters in model training such as learning rate, batch size, and weight for reconstruction loss, we refer to the values which were used in the image to image translation work \cite{isola2017image}.

For training of detection models, i.e., Faster R-CNN ResNet101 and Inception Resnet v2 and R-FCN ResNet101, we use the Tensorflow object detection API \cite{huang2017speed}. The stochastic gradient descent (SGD) method with a momentum of 0.9 is used as an optimizer. In each iteration of the RPN training, 256 training samples are randomly selected from each training image and the ratio between positive (‘polyp’) and negative (‘background’) samples is 1:1. For all other parameters such as learning rate, non-maximum suppression (NMS), and the maximum number of proposals, we use the default values which were used in the original Faster R-CNN work \cite{ren2015faster}.  
Each segmentation model is trained with an Adam optimizer with a learning rate of 0.0001 and batch size of 5 for a different number of epochs to achieve the maximum performance for each model. During training, the training dataset was randomly split into training and validation subsets using 5-fold cross-validation. The 5-fold cross-validation method was performed to choose the best hyper-parameters for the models (e.g. the learning rate relative to the batch size, best epoch, etc) and thus avoid overfitting. The size of the images was changed to 512×512. Finally, we use Jaccard loss combined with a pixel-wise binary cross-entropy to optimize the parameters of the models

\begin{figure*}[!b]
    \centering
    \includegraphics[scale=0.285]{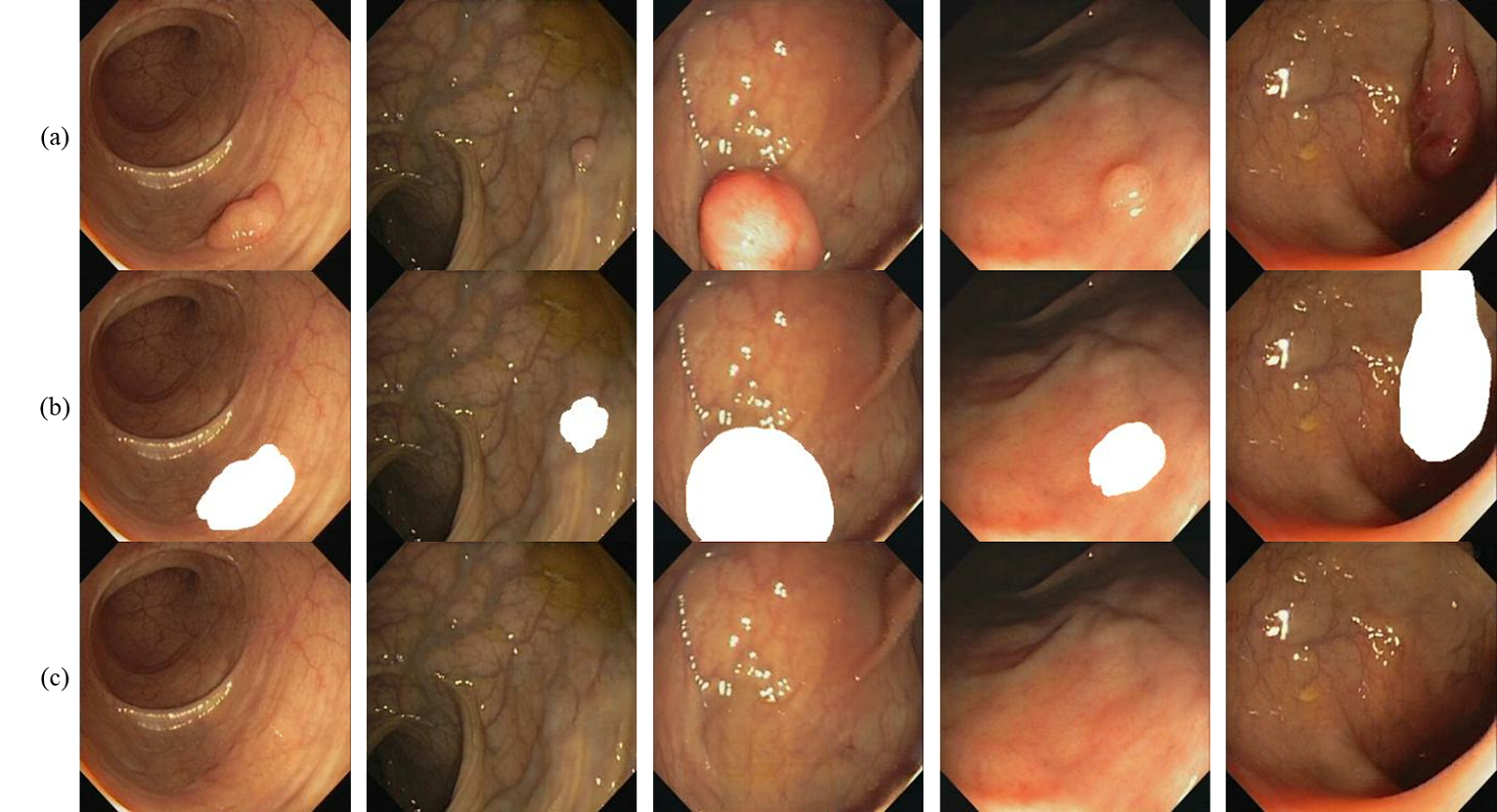}
    \caption{Some results of the polyp to negative translation. Each column represents different image generation result. (a) is polyp image used for preparing the corresponding input conditioned image (b). (c) is generated negative image from the corresponding input condition (b).}
    \label{fig:fig_6}
    \vspace{-4mm}
\end{figure*}

\section{Experimental Datasets}\label{expermint_data}
We use three publicly available polyp datasets, CVC-ClinicDB \cite{bernal2015wm}, ETIS-LARIB \cite{silva2014toward} and CVC-ClinicVideoDB \cite{angermann2017towards}. The CVC-ClinicDB image dataset is used for the whole procedure of the polyp image generation task, i.e., model training for negative and polyp image generation. It is also utilized for training detection and segmentation models as a baseline original image set which is compared to those of the original plus generated synthetic image set. The CVC-ClinicVideoDB and ETIS-LARIB image dataset are used for detection and segmentation model tests, respectively. 

The CVC-ClinicDB image dataset consists of 612 polyp image frames with corresponding 612 binary polyp mask images. The mask images were annotated by skilled clinicians. The 612 polyp image frames have a pixel resolution of 388×284 pixels in SD (standard definition). These images are extracted from 31 colonoscopy videos which contain 34 unique polyps. The CVC-ClinicVideoDB dataset contains 18 SD (384×288 pixels) videos. In this dataset, a total of 10025 frames out of 11954 frames contain a polyp. For each image in the video database, an approximate circle-shaped binary polyp mask is given which makes the dataset suitable for the evaluation of performance improvement the polyp detection models and benchmark our results with the existing studies. For the evaluation of detection models, we use the whole 11954 frames as a test set.

The ETIS-LARIB dataset comprises 196 still images extracted from 34 colonoscopy videos. The images have an HD (high definition) 1255×966 pixels resolution. In this dataset, 44 different polyps are presented in 196 images. A binary polyp segmentation mask annotated by an experienced clinician is provided for each image. This dataset is suitable for the evaluation of performance improvement of the segmentation models because the polyp masks are polygon shapes drawn around the polyp boundaries. 

\vspace{-2mm}
\section{Results and Discussion}\label{results}
\subsection{Generated Negative Images}\label{gen_normal}
In this work, we first transform the given polyp image into a negative image and then create a new polyp in the generated negative image using the polyp condition. In both image translation tasks, i.e., polyp to negative and negative to polyp, we use the same proposed conditional GAN architectures. 

Figure \ref{fig:fig_6} shows some results of the generated negative images from polyp to negative translation task. In each column of Figure \ref{fig:fig_6}, a different generated negative image is represented in (c) which is created from images presented in (b) as the input condition image and (a) is the corresponding original polyp image. Thus, from this task, we transform polyp image (a) to negative image (c). As we can see in Figure \ref{fig:fig_6}, the generated negative images not only clearly remove the corresponding polyp in the polyp portion given by input condition (b), but also naturally harmonize the removed and surrounding parts. We can see that it produces natural negative images even for the third and fifth columns, which are relatively large polyps.
 
\begin{figure}[!ht]
    \centering
    \includegraphics[scale=0.275]{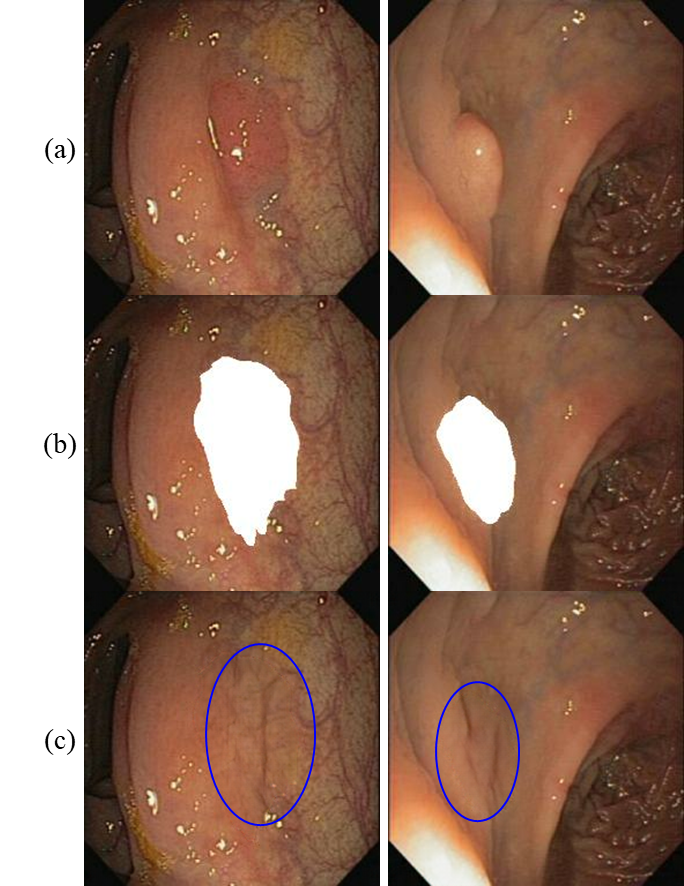}
    \caption{Example of images with large texture changes in the corresponding polyp mask part among the generated negative images.  (a), (b) and (c) represents original polyp image, input condition and generated negative image.}
    \label{fig:fig_7}
    \vspace{-1em}
\end{figure}

Figure \ref{fig:fig_7}  is an example of generated negative images with large texture changes in the part of the binary condition mask. As we can see in the generated negative images (c) of Figure \ref{fig:fig_7}, our trained model adaptively creates a realistic negative colon image by applying the shape and texture of the surrounding parts of the polyp mask. As a result, we can see that capillaries or folds of colon in the generated negative images in the blue circle part of (c).

\begin{figure*}[!b]
    \centering
    \includegraphics[scale=0.28]{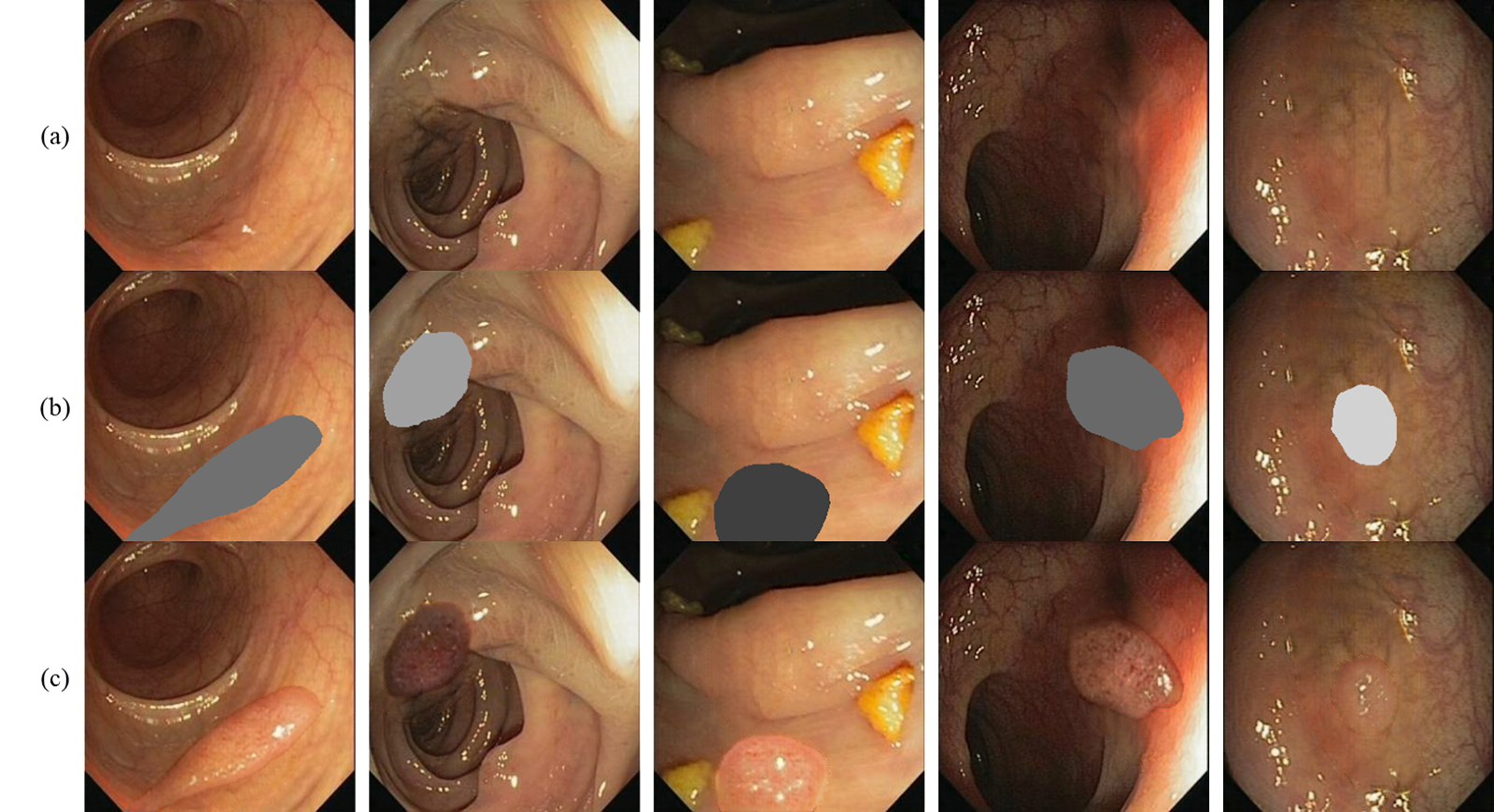}
    \caption{Some results of the negative to polyp translation. Each column represents a different image generation result. (a) is generated negative image from polyp to negative translation. (b) and (c) is the input conditioned image and the corresponding generated polyp image.}
    \label{fig:fig_8}
    \vspace{-1em}
\end{figure*}

\subsection{Generated Polyp Images}\label{gen_polyp}
Our main goal is to create new polyp-looking images. For this purpose, the negative to polyp translation is performed using the previously mentioned method. Figure \ref{fig:fig_8} shows several examples of negative to polyp translation and each column shows a different image result. In this experiment, we use the generated negative image (a) for the preparation of the input condition image (b). Then, our trained model, i.e., negative to polyp model generated output polyp image (c) using the (b) as the input condition.

As mentioned in Section \ref{Normal2Polyp}, in the input condition image (b) of Figure \ref{fig:fig_8}, the value of the polyp mask is randomly selected among 34 different pixel values representing different gray-scale colors. The resulting images in Figure \ref{fig:fig_8} (c) show that the position and shape of the generated polyp are controlled by the polyp mask of (b). In addition, it can be seen that various gray-scale colors and textures are generated rather than similar characteristics, which was a challenge in the existing synthetic polyp generation study \cite{shin2018abnormal}. This is because different pixel values are used as input conditions for the polyp mask part.

\begin{figure*}[h!]
    \centering
    \includegraphics[scale=0.85]{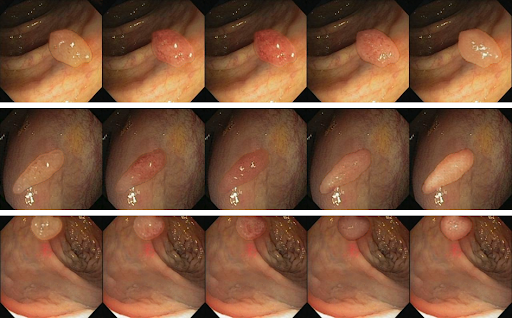}
    \caption{Example of generated polyps. The images in each row are generated when only the pixel value of the polyp mask is changed under the same input condition..}
    \label{fig:fig_9}
    \vspace{-1.0em}
\end{figure*}

To analyze this in detail, we have generated synthetic polyp images by varying the pixel value of the polyp mask part in the same input condition image in Figure \ref{fig:fig_9}. Each row in Figure \ref{fig:fig_9} is an example of a polyp image generation result when only the pixel value of the polyp mask part varies for the same input condition. In other words, each column is generated from different input conditions with different polyp mask values.

As it can be seen from the results of each row in Figure \ref{fig:fig_9}, even with the same shape of polyp mask and background image, it is possible to create polyps with various characteristics in terms of color or texture by controlling the pixel value of the polyp mask. If we look at each column in Figure \ref{fig:fig_9}, we can confirm that the model learns to generate synthetic polyps depending on the polyp mask value rather than the shape or the background image. Because even if different background images and shapes are used, the model can generate polyps with similar characteristics. These two experiments confirm that diverse image generation using the same input, a limitation of the existing image to image transform study \cite{shin2018abnormal, isola2017image}, can easily be achieved by adjusting the mask pixel value of the input condition rather than changing the network structure or training loss. This methodology makes it possible to control the generation of synthetic polyps with characteristics that are difficult for the model to detect, or polyps with clinically malignant characteristics that should not be overlooked. 

Furthermore, we measured the inference speed of the generator network after being trained  within the Conditional GAN framework used in this paper. As a result, the average inference time was measured to be 51.33 msec on an NVIDIA RTX 2080Ti GPU.

\subsection{Evaluation of Polyp Detection Performance}\label{Det_performance}
The advantage of synthetic polyp generation based on the proposed input condition mask is that the generated polyp image can be used directly as a training image for polyp detection or segmentation task without mask annotations. This is significant because it can reduce the manual labeling cost of experts in mask annotation of medical data.

For evaluation of generated polyp images, we first assess polyp detection performance when generated polyp images are used as additional training samples. As polyp detection models, the Faster R-CNN ResNet101, Faster R-CNN Inception ResNet v2, and R-FCN ResNet101 models mentioned in section \ref{Det_Seg} are used. To define evaluation metrics for polyp detection, it is important to specify and compute the following medical terminologies: true positive (TP), false positive (FP), false negative (FN), and true negative (TN) where:

TP = detection output within the polyp ground truth.

FP = any detection output outside the polyp ground truth.

FN = polyp not detected for positive (with polyp) image.

TN = no detection output for negative (without polyp) image.

If there is more than one detection output, only one TP is counted per polyp. Based on the above parameters, we define three performance evaluation metrics, precision (Pre), recall (Rec) and, f1-score (f1):

\vspace{-1.0em}
\begin{multline}
      Pre = \frac{TP}{TP+FP} \times 100. \quad Rec = \frac{TP}{TP+FN} \times 100. \\ f1 = \frac{2 \times {Rec \times Pre}}{Rec \times Pre} \times 100. \quad \quad \quad \quad \quad \quad \quad \quad \quad  
\end{multline} For evaluation of polyp detection performance, the CVC-ClinicVideoDB dataset (see Section \ref{expermint_data} for details) is used as the test dataset.

Table \ref{tab:tab_1} lists the polyp detection performance for three different models. The original columns show the performance when the models are trained with The CVC-ClinicDB image dataset (612 polyp images), which is the originally given training dataset. The paper \cite{shin2018abnormal} columns show the detection results when the models are trained on the original training dataset plus additional 372 synthetic polyp images generated by the method proposed in paper \cite{shin2018abnormal}. The combined columns show the performance when the models are trained by adding 350 polyp images generated by the proposed method in this paper to the original training dataset.

\begin{table}[!ht]
\caption{Comparison of polyp detection performance for the three detection models using the original training set and combined training set by generated polyp image. Number of samples is: Original (612); \cite{shin2018abnormal} (612+372); Combined (612+350) }\label{tab:tab_1}
\vspace{-1em}
\begin{center}
\footnotesize
\begin{tabular}{c c c c c c c c}
\toprule
\rowcolor{Gray}
   \multicolumn{8}{c}{Faster R-CNN ResNet101}  \\
   \cline{1-8}
   \rowcolor{Gray}
   Models & TP & TN & FP & FN & Rec & Pre & F1 \\
   \cline{1-8}
   Original & 6047 & 1431 & 1513 & 3978 & 60.32 & 79.99 & 68.76  \\
   \cite{shin2018abnormal} & 5370 & \textbf{1603} & 1049 & 4655 & 53.57 & 83.66 & 65.31  \\
   Combined & \textbf{6263} & 1508 & \textbf{991} & \textbf{3762} & \textbf{62.47} & \textbf{86.34} & \textbf{72.49} \\
    & & & & & & &    \\
   \rowcolor{Gray}
   \multicolumn{8}{c}{Faster R-CNN Inception ResNet v2} \\
   \cline{1-8}
   \rowcolor{Gray}
   Models & TP & TN & FP & FN & Rec & Pre & F1 \\
   \cline{1-8}
   Original & 6011 & \textbf{1496} & 1333 & 4014 & 60 & 81.9 & 69.22 \\
   \cite{shin2018abnormal} & 6831 & 1399 & \textbf{1177} & 3194 & 68.1 & 85.3 & 75.74 \\
   Combined & \textbf{7056} & 1351 & 1212 & \textbf{2969} & \textbf{70.38} & \textbf{85.34} & \textbf{77.14} \\
    & & & & & & & \\
   \rowcolor{Gray}
   \multicolumn{8}{c}{R-FCN ResNet101} \\
   \cline{1-8}
   \rowcolor{Gray}
   Models & TP & TN & FP & FN & Rec & Pre & F1 \\
   \cline{1-8}
  Original & 5762 & 1304 & 2062 & 4263  & 57.48 & 73.65 & 64.56 \\
  \cite{shin2018abnormal} & 5554 & \textbf{1653} & \textbf{809} & 4471 & 55.4 & \textbf{87.29} & 67.78 \\
  Combined & \textbf{6555} & 1596 & 1032 & \textbf{3470} & \textbf{65.38} & 86.39 & \textbf{74.43} \\
   %& & & & & & & \\
   %\multicolumn{3}{l}{Original: 612 samples} & & & &  & \\ %\multicolumn{3}{l}{\cite{shin2018abnormal}: 612+372 samples} & & & & & \\ %\multicolumn{4}{l}{Combined: 612+350 samples} & & & \\
   \bottomrule
\end{tabular}
\end{center}
\end{table}

The results in Table \ref{tab:tab_1} show that the use of generated images (Combined) demonstrates better polyp detection performance for all three models than the use of the original dataset only. For the three models, both recall and precision increased. This means that polyps missed by the model trained with the existing training dataset are improved when the combined dataset is used. For example, when training using the combined set based on the Faster R-CNN Inception ResNet v2 model, 1045 additional polyps are detected compared to the polyps detected by the original set training. In addition, when the polyp images generated by the proposed method are used for training, improved performance is demonstrated in all three models based on f1-score compared to using the polyp image generated by the existing method \cite{shin2018abnormal}.

In \cite{de2021training}, the same dataset as of this study was used for the evaluation of polyp detection performance, i.e., the detection model was trained on the synthetic polyp generated by their proposed method together with the CVC-ClinicDB training data. They evaluated the detection performance of Faster R-CNN ResNet50 on the CVC-ClinicVideoDB. For comparison, we measure the detection performance for the same model using a dataset generated by our proposed method (i.e., Combined). Table 3 summarizes the detection performance of Faster R-CNN ResNet50 for both methods and shows that our method obtains better results in terms of Precision and F1-score while their model obtained better Recall.   
% We obtain 65.8 recall, 88 precision, and 75.3 f1-score.
% and obtain 73.5 recall, 53.8 precision, and 62.1 f1-score
\begin{table}[!ht]
\caption{Comparison of polyp detection performance of Faster R-CNN ResNet50 between our results and the results shown in \cite{de2021training}}\label{tab:tab_3}
\vspace{-1em}
\begin{center}
\small
\begin{tabular}{c c c c c c c c}
   \toprule
   \multirow{2}{*}{Models} & \multicolumn{6}{c}{Faster R-CNN ResNet50} \\
   \cline{2-8}
    & TP & TN & FP & FN & Rec & Pre & F1 \\
   \cline{1-8}
    \cite{de2021training} & \textbf{7378} & - & 6313 & \textbf{2647} & \textbf{73.5} & 53.8 & 62.1  \\
   Ours & 6597 & 1692 & \textbf{859} & 3428 & 65.8 & \textbf{88} & \textbf{75.3}  \\
   \bottomrule
\end{tabular}
\end{center}
\vspace{-1.5em}
\end{table}

\subsection{Evaluation of Polyp Segmentation Performance}\label{Seg_Performance}
To quantitatively evaluate the effectiveness of the generated synthetic polyps in improving the performance of the segmentation models, we apply commonly used evaluation metrics: Jaccard index (J) also known as intersection over union (IoU) and Dice similarity score (D) as follows:\begin{multline}
    J(A.B) = \frac{\mid A \cap B \mid}{\mid A \cup B\mid} = \frac{\mid A \cap B \mid}{\mid A \mid + \mid B \mid - \mid A \cap B \mid}. \\
     Dice(A.B) = \frac{2\mid A \cap B \mid}{\mid A \mid + \mid B \mid}. \quad \quad \quad \quad \quad \quad \quad \quad \quad \: \: \end{multline} where $A$ refers to a generated output by the models and $B$ is the corresponding ground-truth of $A$. 

Table \ref{tab:tab_2} demonstrates performance evaluation of the three segmentation models where two scenarios in terms of training data, i.e., original and original plus 350 synthetic polyp images (combined training), are used to train the models. As it can be seen from the table, the performance of the three segmentation models improves when the generated synthetic polyps are added to the original training data as follows: The performance of TernausNet-16 improved by 6.44\% Jaccard and 9.66\% Dice, and AlbuNet-34 by 4.11\% Jaccard and 5.87\% Dice, and MDeNetplus by 5.27\% Jaccard,  6.47\% Dice.

\begin{table}[!ht]
\caption{Comparison of polyp segmentation performance for three different models using original training set and combined training set by generated polyp image.}\label{tab:tab_2}
\vspace{-1.0em}
\begin{center}
\footnotesize
\begin{tabular}{c c c c c c c}
\toprule
\multirow{2}{*}{Models} & \multicolumn{2}{c}{TernausNet-16} & \multicolumn{2}{c}{AlbuNet-34} & \multicolumn{2}{c}{MDeNetplus} \\
\cline{2-7}
   & Jaccard & Dice & Jaccard & Dice & Jaccard & Dice  \\ 
 \midrule
 Original & 35.47 & 43.18 & 46.75 & 56.98 & 45.77 & 56.53 \\
 \cite{shin2018abnormal} & 37.86 & 47.57 & 50.48 & 60.31 & 50.49 & 60.77 \\
 Combined & \textbf{41.91} & \textbf{52.84} & \textbf{50.86} & \textbf{62.85} & \textbf{51.04} & \textbf{63} \\
 \bottomrule
\end{tabular}
\end{center}
\end{table}

In Table \ref{tab:tab_3}, we use the same segmentation models to compare the improvement capability of the synthetic polyps generated by this work and the synthetic polyps generated by the method presented in \cite{shin2018abnormal}. For all three models, the synthetic polyps of the current study show a better result improvement than the synthetic polyps of \cite{shin2018abnormal} study. This might be because we are capable of controlling the generation of synthetic polyps with various shapes and features from the same input condition image. As a result, with the proposed method, polyps with various characteristics can be generated. This is a difference from \cite{shin2018abnormal} in which only the shape of the polyp can be controlled.

\subsection{Limitations and Future Work}
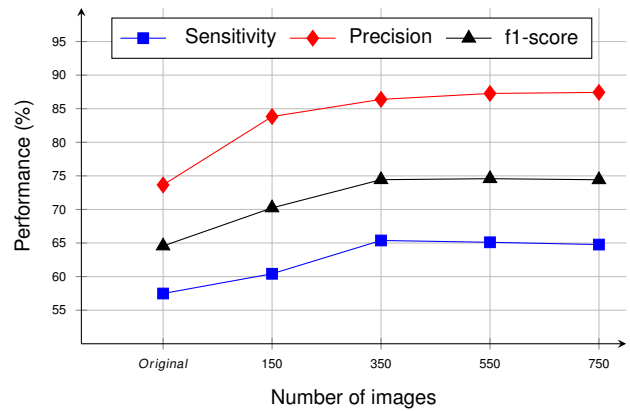
\begin{figure}[!h]
\centering
\begin{tikzpicture}[font=\footnotesize\sansmath\sffamily]
\begin{axis}[
      %legend pos=south center,
      legend style={at={(0.055,0.91)}, legend columns=-1, anchor=west, style={column sep=0.1cm}},
      domain = 0:100,
      axis on top=false,
      axis x line=middle,
      axis y line=middle,
      xlabel     =  Number of images,
      xlabel near ticks,
      ylabel     =  Performance (\%),
      ylabel near ticks,
      width=8.75cm,
      height=6cm,
      clip       = true, 
      xmin = 0,  xmax = 1000,
      ymin = 50, ymax = 100,
      grid=major,
      xtick={150,350,550,750,950},
      xticklabels={$Original$,$150$,$350$,$550$,$750$},
      ytick={55,60,65,70,75,80,85,90,95},
      yticklabels={$55$,$60$,$65$,$70$,$75$,$80$,$85$,$90$,$95$}]
      
    \addplot[
    color=blue,
    mark=square*,
    ]
    coordinates {
    (150,57.48)(350,60.42)(550,65.38)(750,65.11)(950,64.77)
    };
    \addplot[
    color=red,
    mark=diamond*,
    mark size=2.9pt,
    ]
    coordinates {
    (150,73.65)(350,83.83)(550,86.39)(750,87.27)(950,87.43)
    };
    
    \addplot[
    color=black,
    mark=triangle*,
    mark size=2.9pt,
    ]
    coordinates {
    (150,64.56)(350,70.23)(550,74.43)(750,74.58)(950,74.42)
    };
    \legend{\;Sensitivity,Precision, f1-score}
\end{axis}
\end{tikzpicture}\vspace{-1mm}
\caption{Performance improvement with different number of synthetic polyp images added to the training data}
\label{fig:fig_10}
\vspace{-1em}
\end{figure}

One natural question might be how performance changes as the number of generated polyps increases. In Figure \ref{fig:fig_10}, the detection performance of the RFCN ResNet101 model is evaluated when different numbers of synthetic polyp images are added to the original training dataset. As it can be seen that the model starts gradually improving its performance when it is exposed to additional synthetic polyp images. However, the model reaches a saturation performance improvement after a certain point even if more synthetic polyp images are added to the training set. This is due to the limitation of this method that is unable to introduce new unseen features. This method only manipulates the existing features in the training dataset used to train the GAN model and tries to reuse the same set of features to generate new-looking synthetic polyps. In other words, the proposed framework cannot improve data distribution and is unable to add new features to the training data, it only leads to an image-level transformation and feature-level manipulation. This is a common limitation of GAN-based image generation models.

%\hl{In the last experiment, we trained AlbuNet34 on 612 generated synthetic polyps only. The model obtained 39.98\% of Jaccard index and 48.03\% of Dice similarity score. This result shows that the generated synthetic polyps carry polyp information and can still be used to train a model. However, this result is not as good as the result obtained by the original polyp images. This is due to the degraded quality of the generated polyp images compared to the high quality of the original images.} 

For future work, studies that generate polyps with new characteristics not included in the training data will be needed. Performance can further be improved if characteristics of two/multiple polyps can be fused into a single synthetic polyp. This could be done by fusing obtained abstract features from different polyps or by manipulating the features by adding random noise. Both approaches could be performed in the latent space. Alternatively, we can apply a method such as CAN (Creative Adversarial Networks) \cite{elgammal2017can} that adjusts the loss term of GAN to generate a creative image by maximizing deviation from established styles and minimizing deviation from the distribution of several datasets.

\section{Conclusion}\label{conclusion}
It is expensive to acquire images of various types of polyps that can be used for training deep learning based automatic polyp detection and segmentation models. We proposed a synthetic polyp image generation framework based on the conditional GAN architectures. In the proposed framework, we converted a given polyp image into a negative image and then the negative image back into a new-looking synthetic polyp image using the same networks for both tasks. In the previous polyp generation studies, a single value was assigned to the polyp mask in the condition input image. This led the models to face difficulties to generate various polyps with different characteristics. In this study, we attempted to overcome this shortcoming. We developed a framework to generate various polyps with different features by controlling the value of the polyp masks in the input condition images. The experimental results showed that the proposed framework could generate polyps with various characteristics similar to the real ones. In addition, it was confirmed that the polyp detection and segmentation performance could be improved when the generated synthetic polyp images were used to train several detection and segmentation models.

\section*{Acknowledgements}
This research was supported by the Research Funds of Mokpo National University in 2020. The authors would like to thank Jacob Bergsland at Oslo University Hospital for his valuable comments.

%%Harvard
%\bibliographystyle{model2-names.bst}\biboptions{authoryear}

\bibliographystyle{elsarticle-num}\biboptions{sort&compress}
\bibliography{refs}

\end{document}